\documentclass{nature}

\usepackage{graphicx}

\title{HCN ice in Titan's high-altitude southern polar cloud}

\author{Remco J. de Kok$^{1,2}$, Nicholas A. Teanby$^{3}$, Luca Maltagliati$^{4}$, Patrick G. J. Irwin$^{5}$, \& Sandrine Vinatier$^{4}$}

\begin{document}

\maketitle

\begin{affiliations}
\item Leiden Observatory, Leiden University, Postbus 9513, 2300 RA, Netherlands
\item SRON Netherlands Institute for Space Research, Sorbonnelaan 2, 3584 CA Utrecht, Netherlands
\item School of Earth Sciences, University of Bristol, Wills Memorial Building, Queen's Road, Bristol, BS8 1RJ, UK
\item LESIA-Observatoire de Paris, CNRS, UPMC Univ. Paris 06, Univ. Paris-Diderot, France
\item Atmospheric, Oceanic \& Planetary Physics, Department of Physics, University of Oxford, Clarendon Laboratory, Parks Road, Oxford, OX1 3PU. UK
\end{affiliations}

\begin{abstract}
Titan's middle atmosphere is currently experiencing a rapid change of season after northern spring arrived in 2009\cite{tea12,vin13}. A large cloud was observed for the first time above Titan's southern pole in May 2012, at an altitude of 300 km\cite{wes13}. This altitude previously showed a temperature maximum and condensation was not expected for any of Titan's atmospheric gases. Here we show that this cloud is composed of micron-sized hydrogen cyanide (HCN) ice particles. The presence of HCN particles at this altitude, together with new temperature determinations from mid-infrared observations, indicate a very dramatic cooling of Titan's atmosphere inside the winter polar vortex in early 2012. Such a cooling is completely contrary to previously measured high-altitude warming in the polar vortex\cite{tea12}, and temperatures are a hundred degrees colder than predicted by circulation models\cite{ran05}. Besides elucidating the nature of Titan's mysterious polar cloud, these results thus show that post-equinox cooling at the winter pole is much more efficient than previously thought.

\end{abstract}

In May 2012, a large cloud-like structure was identified above Titan's dark southern pole by Cassini's Imaging Science Subsystem (ISS)\cite{wes13}. Ever since, it has been seen at very high altitudes ($\sim$ 300 km) and high southern latitudes, at all visible and near-infrared wavelengths. Clouds require temperatures cold enough for atmospheric gases to reach saturation. Hence, clouds on Titan have previously been found near the tropopause and lower stratosphere, where the atmosphere is coldest\cite{gri98,sam97b,and10,gri06}. Instead of a temperature minimum, a temperature maximum was present before 2012 at the altitudes and latitudes where the high-altitude ISS cloud is seen\cite{ach11,tea12}. Such high temperatures precluded the condensation of any of Titan's known trace gases. The presence of a cloud at this location is therefore highly unexpected.

We analysed near-infrared spectra of the high-altitude cloud from Cassini's Visual and Infrared Mapping Spectrometer (VIMS) to constrain its composition and optical thickness. Near-infrared wavelengths are sensitive to vibrational bands of solids and liquids and can therefore be used to uniquely identify the cloud composition. We have averaged the spectra of the high-altitude cloud of 13 similar VIMS image cubes from 29 November 2012, with pixel scales between 89 and 135 km (Fig.~1a), to obtain a high signal-to-noise near-infrared reflectance spectrum of the cloud. This spectrum indeed shows two large spectral features (Fig.~1b), which are not present in cloudless regions. The spectral features coincide exactly with the features expected from HCN ice\cite{del96,moo10} and are detected at a level at least 15 times the standard deviation of reflectance of a single pixel in a single image. Other possible condensates clearly do not match the spectral features seen in the data. We fitted the reflectance spectrum of the cloud and found an excellent agreement with a simple model that includes scattering by an optically thin cloud that is composed of HCN ice particles with a radius between 0.6-1.2 $\mu$m (Fig.~1b). Besides an earlier tentative identification of HCN ice\cite{sam07}, and indirect evidence\cite{lav11}, this is the first time there is strong evidence for HCN condensation in Titan's stratosphere. 

We obtained an estimate of the optical thickness of the cloud from a VIMS image cube from 7 June 2012, which has a pixel scale of 91 km, where the cloud was seen at the limb of Titan (Fig.~2). We determined the cloud top to be located at 300$\pm$70 km, based on the fact the highest blue cloud pixels intersect the 300~km line through the middle. This estimate is consistent with results from the ISS instrument\cite{wes13}, but has a greater uncertainty due to VIMS' larger pixel size. In Fig.~2, sunlight passes a slant path through the atmosphere, before being scattered back by the cloud in almost the same direction to the Cassini spacecraft.  In this geometry the reflectance spectrum is dominated by the reflecting properties of the cloud and the transmission of the atmosphere across the slant path. Unlike in Fig.~1a, there is little background contribution from lower altitudes, making it easier to assess the optical thickness of the cloud. At 2.7 $\mu$m the atmosphere is expected to be practically transparent at 300 km, even for slant paths\cite{bel09,mal14} and almost all signal will be caused by scattering of the cloud. Using the single-scattering approximation for low optical thicknesses, the reflectance can be assumed to be the product of the slant optical thickness, the single-scattering albedo of the particle, and the phase function at the scattering angle, divided by four. The measured reflectance at 2.7 $\mu$m of 0.0028$\pm$0.0002 then directly relates to a slant optical thickness of $\sim$0.09$\pm$0.006, assuming micron-sized HCN particles. Since the single-scattering albedo and phase function at 2.7$\mu$m do not change rapidly with particle size, the slant optical thickness is accurate within a factor of two for the particle range 0.6-1.2 $\mu$m. If the vertical extend of the cloud is ten times smaller than the slant path, this translates to a vertical optical thickness between 0.01-0.07 for particles between 0.6-1.2 $\mu$m at a wavelength of 0.9 $\mu$m, which is a wavelength that can be compared to analysis from ISS measurements.  Although the slant optical thickness can be measured relatively well, an estimate of the particle density requires knowledge of the path length through the cloud, and the exact pressure of the cloud. We perform an order-of-magnitude estimate here using conservative errors on the pressure and path length.  Assuming a slant path length of tens to hundreds of kilometers (depending on the three-dimensional extend of the cloud) and a pressure of 0.1-0.5 mbar (corresponding to altitudes between 200-300 km), this slant optical thickness translates to a particle density of the order of 10$^{4}$-10$^{5}$ particles per gram of gas for micron-sized particles. This particle density falls well within the expected particle density range of micron-sized HCN particles with a reasonable downward wind speed of the order of 0.1-1 mm/s\cite{dek08}. This wind speed is of the same order as the previously inferred downward velocities at the South Pole during southern autumn\cite{tea12,vin13}. Hence, the optical thickness of the cloud is within the range of that expected for HCN particles, if the temperature is cold enough for them to exist.

At the same time as the appearance of the cloud in ISS and VIMS observations, a condensate feature also appeared at the South Pole in far-infrared spectra from Cassini's Composite InfraRed Spectrometer (CIRS)\cite{jen12b}. This feature had been previously observed at the northern pole by Voyager\cite{cou99} and by Cassini since its arrival in 2004\cite{dek07b,sam07,and11}. It could potentially be linked to the condensation of HCN, since it appeared at a location where HCN condensation was expected to give a large cloud signature\cite{dek08}. Unfortunately, the available CIRS observations to date cannot constrain the altitude of the far-infrared condensate feature in the south, making it impossible to firmly establish a connection between this feature and the HCN cloud discussed in this paper. A limb scan with high spatial resolution, which will resolve this issue, is not planned until at least 2015 due to the orbital geometry of Cassini.

Although the VIMS observations of the high-altitude cloud are entirely consistent with the presence of HCN ice particles, the required cold temperatures of $\sim$125 K are unexpected. The best way to study Titan's middle atmospheric temperatures is to use mid-infrared spectra from CIRS. Assuming methane is uniformly mixed in the stratosphere, temperatures can be derived from emission of the $\nu_4$ methane band between 7-8 $\mu$m. CIRS measurements of Titan's limb can derive spatially resolved temperature-pressure profile up to an altitude of at least 400 km\cite{tea12,vin13}. In February 2012, CIRS observations showed an unexpected temperature decrease of the mesosphere by about 35 K compared to one year earlier\cite{vin13}. Unfortunately, no CIRS limb measurements at the location of the high-altitude polar cloud exist after its appearance in May 2012. However, a set of spectra is available from 14 October 2013 that looks down on the South Pole. These spectra probe temperatures at a limited range of altitudes only, but they can be used to determine whether further cooling has occurred at the South Pole after February 2012. Retrievals of temperatures on 14 October 2013, using the NEMESIS retrieval code\cite{irw08,tea10}, are plotted in Fig.~3, which shows that Titan's stratosphere has cooled significantly below 200 km after 2011. Mid-infrared limb measurements of the South Pole are again only planned from 2015 onwards.

The main cooling mechanism in Titan's stratosphere is thought to be radiative cooling\cite{tom08heat}, and a decrease of solar irradiation at the South Pole after equinox can give rise to a cold winter pole. The cooling time scale becomes shorter at greater altitudes, so it is plausible that at 300 km temperatures have dropped even more than below 200 km, especially since a very strong increase in trace gas concentrations has been observed since 2011\cite{tea12,vin13}. These gases radiate strongly in the infrared and hence can produce a strong cooling. On the other hand, adiabatic heating due to the sinking motion of the air at the South Pole has been observed in 2011 around 300 km\cite{tea12,vin13}, so the overall temperature is affected by a combination of chemistry, dynamics and insolation. The LMD global circulation model, which couples the effects of dynamics, haze formation and chemistry\cite{ran05,leb12}, predicts a large temperature maximum around 300 km, which increases in temperature between 2012 and 2013 (Fig.~3). Our detection of HCN ice particles at these altitudes indicates that the polar atmosphere is roughly 100 K colder than predicted at these altitudes, and thus requires the radiative cooling to be far stronger than the adiabatic heating, contrary to expectations. Hence, models of Titan's circulation require revision to understand the transitional behaviour of Titan's atmosphere around equinox.

\textbf{Methods}
We calculated the cloud spectrum of Fig.~1b by taking the mean of the pixels containing the cloud, and combining these for 13 similar, consecutive images (v1732906961-v1732924296), weighted by their standard deviation squared. The VIMS limb spectrum is the mean from cloud pixels in Fig.~2 (image cube v1717755608).  We fitted the cloud spectrum of Fig.~1b to qualitatively demonstrate the presence of HCN particles, to look for other condensates, and to obtain an estimate for the particle size. The spectrum is sensitive to the particle size, with small particles giving a stronger blue slope of the extinction cross-section, and large particles also having less pronounced absorption features. We assumed the spectrum consists of two components: a low-altitude component and a cloud component. The low-altitude component was obtained by taking the mean in two areas of roughly 5x5 pixels on either side of the cloud, at similar solar incidence angles. Spectra from the 13 images were combined as for the cloud spectrum. The background was scaled in the fitting procedure, with the exception of the non-LTE emission at 3.3 $\mu$m. The non-LTE emission was found to be strongly reduced in the cloud spectrum, which could be indicative of cold temperatures. The cloud component consisted of the scattering cross-section of HCN particles, multiplied by its phase function at the scattering angle of the cloud (both calculated by Mie theory using refractive indices at 120 K\cite{moo10}). Furthermore, this cross-section was multiplied by the atmospheric transmission through a slant path at 250 km, as measured by VIMS\cite{mal14}. Free parameters were the particle size, a scaling factor for the transmission spectrum, and a scaling factor for the low-altitude component, after which the cloud component was made to fit the observed spectrum between 3.6-4.0 $\mu$m. These parameters were explored along a wide grid and for each particle size, the best fit to the spectrum was evaluated. Particle sizes between 0.6-1.2 $\mu$m gave qualitatively good fits to the observed spectrum. Smaller particles could not reproduce the general slope of the observed spectrum, and overstimated the 3.2 $\mu$m feature compared to the 4.8 $\mu$m feature. Large particles could not reproduce the spectral slope and had less pronounced absorption features in the best fit. Note that a more quantitative fit of the observations would require a three-dimensional radiative transfer model, to better model the atmospheric transmission and the low-altitude contribution near the terminator.

Temperatures are obtained from an average of 0.5 cm$^{-1}$ resolution CIRS spectra from 14 October 2013, all within 5$^\circ$ of the South Pole. We performed retrievals on the $\nu_4$ emission band of CH$_4$, covering the spectra range 1240--1360~cm$^{-1}$, using the NEMESIS retrieval code\cite{irw08}. We assumed a  CH$_4$ volume mixing ratio of 1.48\% and previously derived haze spectral properties\cite{vin12}. We refer for more details on the retrieval procedure to a previous paper\cite{tea10}.



\begin{addendum}
\item RJdK thanks the PEPSci program of the Netherlands Organisation for Scientific Research (NWO). NAT and PGJI thank the UK Science and Technology Facilities Council. LM thanks the Agence Nationale de la Recherche (ANR Project``APOSTIC" no11BS56002, 968 France). We thank B.~B\'{e}zard, T.M.~Ansty, C.~Nixon and  M.~L\'{o}pez-Puertas for discussions, and the referees for their comments. We thank the VIMS and CIRS operation and calibration teams.
\item[Author Contributions] RJdK conceived the study. RJdK, LM, NAT, and PGJI performed the VIMS analysis. NAT and SV performed the CIRS analysis. All authors contributed to the interpretation, in addition to editing and improving the final manuscript. 
\item[Author Information] Reprints and permissions information is available at www.nature.com/nature. 
The authors declare that they have no competing financial interests.
Correspondence and requests for materials should be addressed to RJdK. (email: R.J.de.Kok@sron.nl).
\end{addendum}



\clearpage
\begin{figure}
\includegraphics[width=150mm]{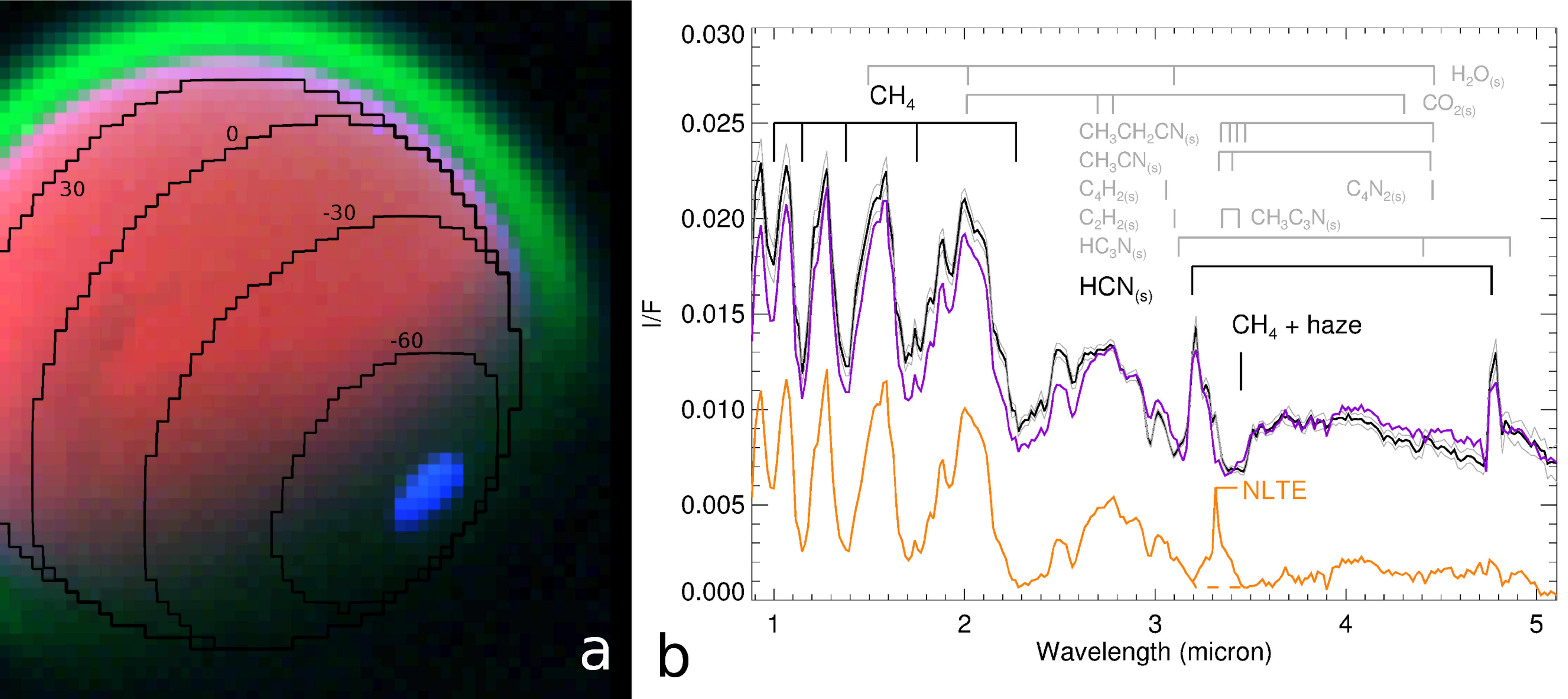}
\caption{
\noindent
{\bf Identification of HCN ice in VIMS observations.}
(a) A single false-colour VIMS image from 29 November 2012 indicating the illuminated surface (1.07 $\mu$m, red), non-LTE emission (3.33 $\mu$m, green), and an HCN ice feature (3.21 $\mu$m, blue). Contours show surface latitudes. Solar illumination is from the upper left. (b) Mean spectrum away from the cloud (orange, dashes indicate version without non-LTE) and within the cloud (black, offset by 0.005 for clarity, with standard deviation from a single pixel). Wavelengths of HCN ice features, and of features from other possible condensates\cite{del96,moo10,war86,war08} are indicated. A fit to the cloud spectrum is plotted in purple. 
}
\end{figure}

\begin{figure}
\includegraphics[width=90mm]{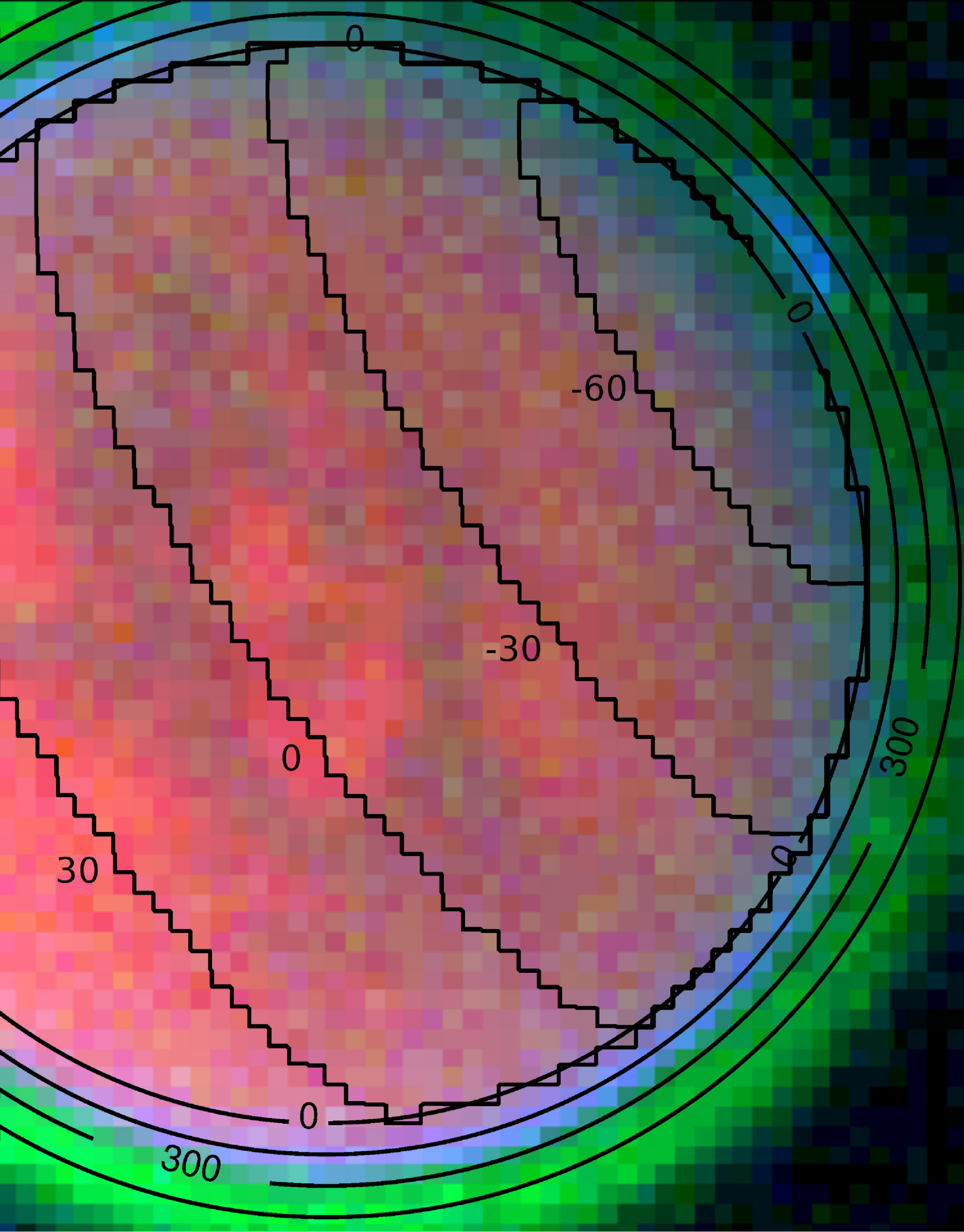}
\caption{
\noindent
{\bf Cloud observed at Titan's limb.}
False-colour VIMS image from 7 June 2012 showing Titan's polar cloud at the limb, with altitudes (km) and surface latitudes indicated. Colours are as in Fig.~1a. Illumination is from behind the observer. The cloud is seen at the top right. The blue/purple color of the entire limb is caused by Titan's visible disc being larger at 3.2 $\mu$m compared to 1.07 $\mu$m. In this image, the cloud is relatively weak compared to the rest of the disc, making the limb relatively more blue than in Fig.~1a.
}
\end{figure}

\begin{figure}
\includegraphics[width=150mm]{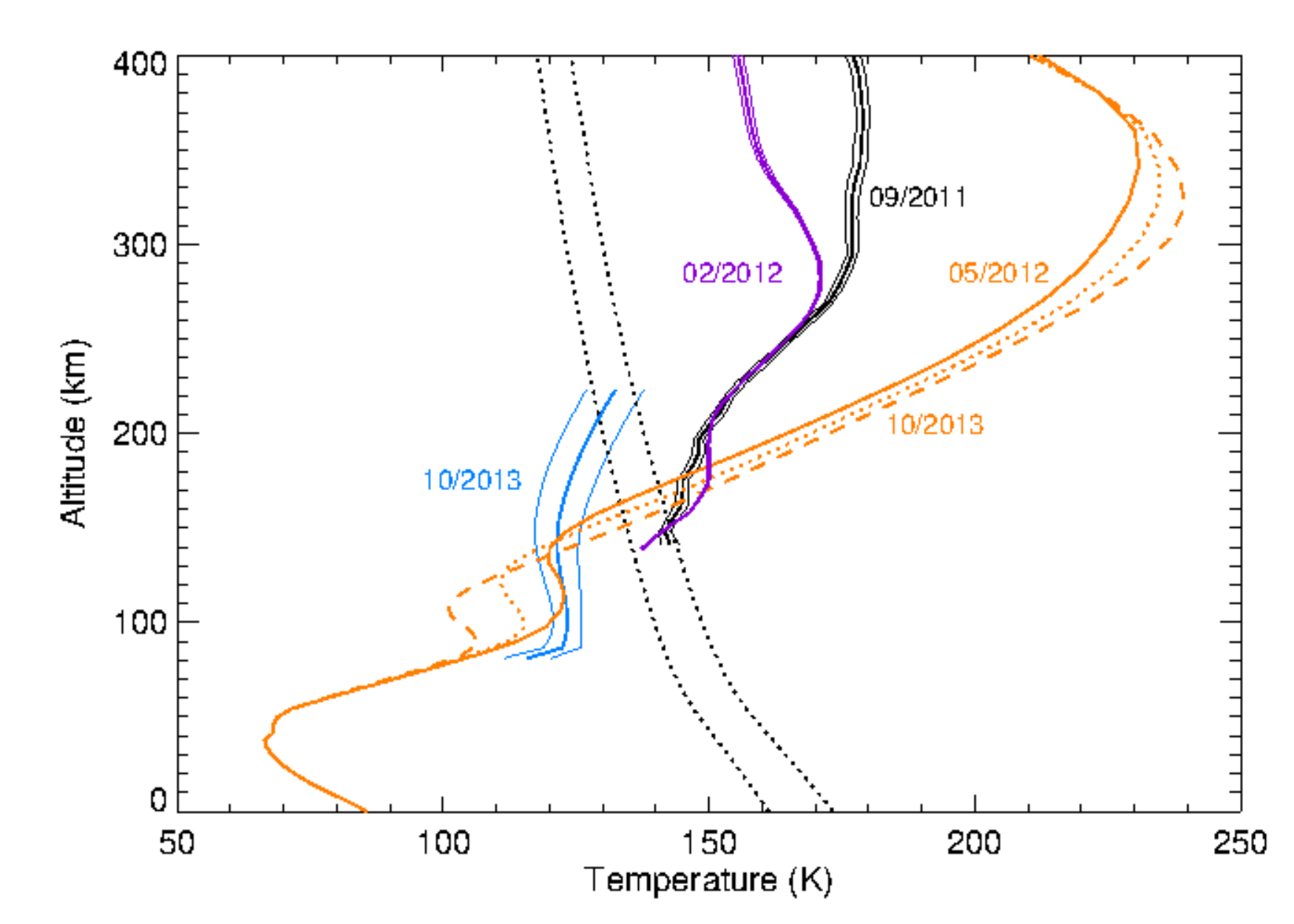}
\caption{
\noindent
{\bf South-polar temperatures from models and retrievals.} 
Retrieved temperatures and their 1$\sigma$ errors at 86$^\circ$S in September 2011\cite{tea12} (black) and February 2012\cite{vin13} (purple) from CIRS limb measurements  and at 87$^\circ$S in October 2013 from CIRS nadir measurements (blue). We plot only regions where the observations provide reliable temperature information. Orange lines are circulation model output\cite{ran05} for May 2012 (solid), December 2012 (dotted), and October 2013 (dashed). Black dotted lines indicate saturation temperatures for HCN volume mixing ratios of 10$^{-6}$ (left) and 10$^{-5}$ (right), which cover the measured concentrations of HCN in the south polar vortex\cite{tea12,vin13}. A cloud at 300 km would require temperatures of $\sim$125 K there.
}
\end{figure}


\end{document}